\documentclass[12pt, journal, a4paper]{IEEEtran}
\usepackage{graphicx} 
\usepackage{subcaption}
\usepackage{tabularx}
\usepackage{tabulary}
\usepackage{url}       
\usepackage{amsmath} 
\usepackage[left=2cm,right=1.5cm,top=1.5cm,bottom=1.5cm]{geometry}
\usepackage{amsfonts}
\usepackage{amssymb}
\usepackage{textcomp,gensymb}
\usepackage{gensymb}
\usepackage{float}
\usepackage{caption}
\usepackage[table,xcdraw]{xcolor}
\usepackage{authblk}
\usepackage{siunitx}
\usepackage{afterpage}
\usepackage{placeins}
\usepackage{wrapfig}
\usepackage{tabularx} % Include this in your preamble if you use tabularx
\usepackage{wrapfig} 
\usepackage{booktabs}
\title{Classification of Spam URLs Using Machine Learning Approachs \vspace{1em}}

\author[1]{Omar Husni Odeh}
\author[2]{Anas Arram}
\author[2]{Murad Njoum}
\affil[1]{An-Najah National University}
\affil[2]{Department of Computer Science, Birzeit University}

\date{April 2021}
\providecommand{\keywords}[1]
{
  \textbf{\textit{Keywords---}} #1
}

\begin{document}

\maketitle
\begin{abstract}
The Internet is used by billions of users every day because it offers fast and free communication tools and platforms. Nevertheless, with this significant increase in usage, huge amounts of spam are generated every second, which wastes internet resources and, more importantly, users' time.  This study investigates the use of machine learning models to classify URLs as spam or non-spam. We first extract the features from the URL as it has only one feature, and then we compare the performance of several models, including k-nearest neighbors, bagging, random forest, logistic regression, and others. Experimental results demonstrate that bagging outperformed other models and achieved the highest accuracy of 98.64\%. In addition, bagging outperformed the current state-of-the-art approaches which emphasize its effectiveness in addressing spam-related challenges on the Internet. This suggests that bagging is a promising approach for URL spam classification.
\end{abstract}
\section{Introduction}
The Internet is an open space for everyone to freely create content, publish it, and share it with others. In the last decade, internet access has increased tremendously. This increase in audience came with some side effects. Too many ads and spam are being shared everywhere, whether it is email or any other type of social media. Platforms like email clients which are used by hundreds of millions of users every day struggle to effectively filter the content to the end user, \cite{1}\cite{2}.
\newline

Blacklist is one of the common methods used to identify malicious URLs. Although blacklisting has been effective for many URLs, the rapid increase in these URLs makes it an insufficient method. Therefore, Machine learning techniques have been proposed to address this issue \cite{3}. These techniques can detect malicious websites, even if they have never been encountered before. In the context of this discussion, machine learning (ML) refers to the field of computer algorithms that autonomously enhance their performance through experiential learning and data analysis \cite{4,22}. Because ML models possess the capability to comprehend the underlying structural patterns within URLs, they provide more insightful methods for classifying URLs \cite{5,6,7}.
\newline

In this study, we want to investigate the use of machine learning models to classify URLs that are most likely spam. The model's input and output are pretty straightforward; it will take a URL and it will classify it as spam or not. Since we are taking only one feature as an input, we will need to analyze and extend this feature to extract more information about the URL to determine its characteristics, \cite{8, 21}. In addition, we will build multiple machine learning models, with different parameters to evaluate different options, and finally to come up with the best model that can result in the highest possible accuracy. 
\newline

The paper is organized into five sections. Section 2 covers related works in spam URLs. Section 3 presents the proposed ML models, while Section 4 discusses the obtained results. Finally section 5
concludes the paper. 
\section{Literature review}
URL spam detection is a modernistic field that form a solid interest for both organizations and researchers. It started to receive more and more attention from researchers due to the major evolving and exposing that happened on the internet over the past few years. Previous work on this topic has involved analysis of the URL and the page itself.
\newline

Oshingbesan \emph{et al}, 2023, \cite{18} examined different ML models to classify malicious websites across different datasets. From their results, K-Nearest neighbor performs the best in classifying malicious URLs. One the otherside, other models like random forest, decision trees, logistic regression, and support victor machine, outperform the baseline models across all dataset. 
\newline

Murat Koca \emph{et al}, 2022, \cite{20} investigated the use of different ML models for classifying the URLs. These models include Logistic Regression, Neural Networks, and multiple Naive Bayes Algorithms. The results showed that the Naive Bayes model performed noticeably better than both the logistic regression and neural network approaches across all tested dataset. 
\newline

Gyongyi and Garcia-Molina, 2005, \cite{8}, attempted to classify the web spam into smaller buckets, such as the URL spam, redirection, and keywords stuffed in the link. While splitting and categorizing the spam into some specific buckets will likely improve the classifier ability to detect spams, their paper focused on building a general classifier for all different types of spam.
\newline

Ntoulas, Najork, Manasse, and Fetterly, 2006, \cite{9}, studied and analyzed the content of the page itself, and it typically included creating and extracting features from the HTML structure of the page, JavaScript, and links, such as the number of the words on the page, and the average length of words, and the number of words in both title and body. Other feature extraction methods involved looking at the percentage of hidden content, which is not visible to the user who is browsing a specific page.
\newline

Egele, Kolbitsch, and Platzer, 2009, \cite{10}, had another approach which starts by determining the important features in terms of their rank in a search engine and then find the features that are most likely to be used by spammers. The problem of this approach is that it is infeasible to enumerate all ranking element, and thus, some important features may be missed.
\newline

Boser, Guyon, and Vapnik 1992, \cite{11}, said that all the models are based on SVMs, which is known to perform well in classification tasks, Joachims 1998, \cite{12,23}. Their evaluation used the standard area under ROC curve metric over a K-fold cross-validation where K=10, and they used some tools provided by libsvm Chang and Lin, 2001, \cite{13}. The feature selection was made using frequency count over an entire sample set. All their charts were plotted only for feature sizes of < 1000. Larger feature sizes did not significantly improve results.
\section{Dataset and features}
\subsection{Data description}
To train our models, we are using URL - Spam or Not Spam - classification dataset, Dec. 2021, \cite{14}. The dataset contains about 148.3K URLs in which one-third are flagged as a spam URL and the rest are not spam. It can be used to create a binary classification model. The dataset was created by the pudding, \cite{15}. The dataset links were found in different newsletters. Their flagging system identifies if a link is a spam or not, by parsing links from over 100 newsletters every 30 minutes. A link is programmatically flagged if it appears more than three times in a single newsletter or contains a likely subscribe/unsubscribe URL.

\subsection{Features Extraction}
The dataset has only one input feature which is the URL itself. To make the best use of these URL we need to extract as many features as we can to understand the characteristics of the URL in order to identify some patterns which can be useful for the machine learning models later on.
Below is the list of most important extracted features with a brief description:

\begin{table}[!htbp]
\caption{Dataset new features}
\small
\begin{tabular}{|l|l|}
\hline
\rowcolor[HTML]{FFFE65} 
Feature   name  & Description                            \\ \hline
url\_length     & The number of characters.              \\ \hline
has\_subscribe  & Is the URL has the word subscribe.     \\ \hline
contains\_hash  & Whether the URL has the hash letter.   \\ \hline
num\_digits     & The number of digits in the URL.       \\ \hline
non\_https      & Is it a secure connection.             \\ \hline
num\_words      & Number of words in the URL.            \\ \hline
entropy         & The measure of disorder/uncertainty.   \\ \hline
num\_params     & The number of query parameters.        \\ \hline
num\_fragments  & Number of fragments in the URL.        \\ \hline
num\_subdomains & The number of sub domains.             \\ \hline
num\_\%20       & Number of encoded white spaces         \\ \hline
num\_@          & Number of @ in the URL.                \\ \hline
has\_ip         & Whether it's a FQDN or IP address.     \\ \hline
\end{tabular}
\end{table}

\subsection{Exploratory data analysis}
\begin{figure}[ht]
\centering
\captionsetup{justification=centering}
  \includegraphics[scale=0.6]{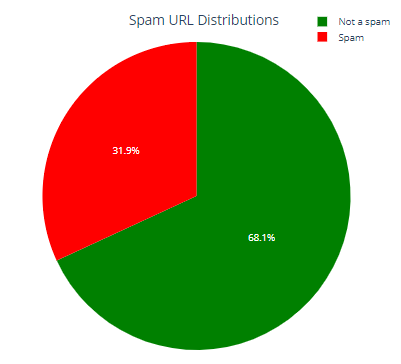}
  \caption{Spam URL distributions}
  \label{fig:mesh1}
\end{figure}

Two-thirds of the used dataset are not spam URLs, which is more than 100k URLs. The spam URL represents ~32\% of the data, which is about 48k URLs. Figure \ref{fig:mesh1} represents the distribution of spam vs. non-spam URLs.

The URL length analysis shows that most of the spam URLs have a length of less than 100 characters as shown in the histogram below:

\begin{figure}[ht]
\centering
\captionsetup{justification=centering}
  \includegraphics[scale=0.4]{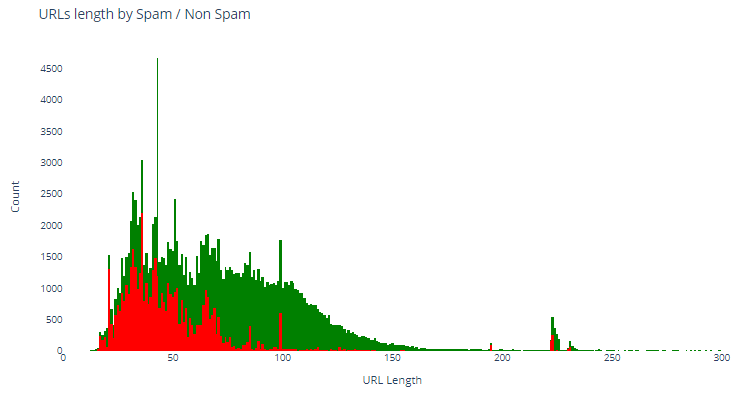}
  \caption{URLs length by Spam / Non spam}
\end{figure}

\begin{figure}[ht]
\centering
\captionsetup{justification=centering}
  \includegraphics[scale=0.6]{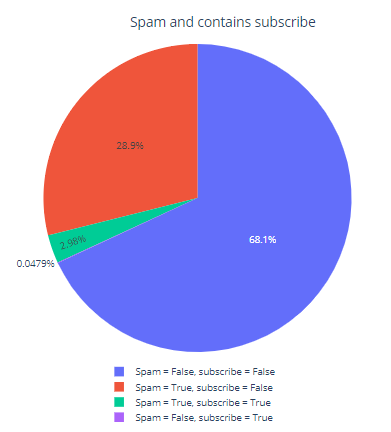}
  \caption{URL has the word subscribe}
  \label{fig:words}
\end{figure}

Another important piece of information extracted from these URLs is that the URLs that contain subscribe words are most likely spam. There is 3\% of the input URLs have the word subscribe and almost all of them are spam. 
In addition, the number of words in the spam URLs is less than 5 in the majority of the data, unlike the non spam which is distributed over a wider range as shown in Figure \ref{fig:words}.

\begin{figure}[ht]
\centering
\captionsetup{justification=centering}
  \includegraphics[scale=0.6]{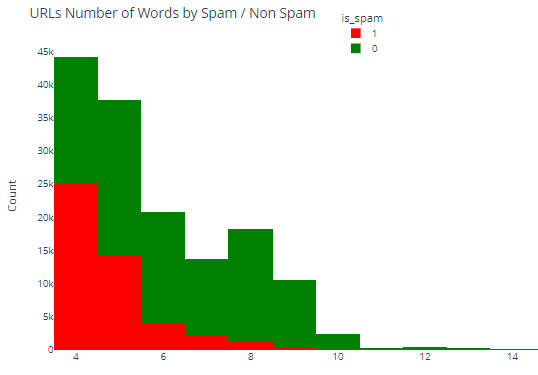}
  \caption{URLs number of words by Spam / Non spam}
  \label{fig:spams}
\end{figure}

There is about 2.08\% of the URLs that are not using secure HTTPS protocol, 1.25\% of these URLs are spam which is more than 60\% of the HTTP URLs knowing that only 33\% of the total data is flagged as spam. This indicates a usable feature for the model as shown in \ref{fig:spams}.

\section{Methods}
The following subsections describe the machine learning models used in this study to classify spam URLs.
\subsection{Logistic regression}

Even though it has the word regression, it's a classification model rather than a regression model. It's a simple and efficient method especially for binary and linear classification problems. It is a model, which that's very easy to realize, and it achieves a very good performance with linearly classes. It's an extensively algorithm for classification. The logistic regression model is a statistical method for binary classification which can be also generalized to multiclass classification. Scikit-learn has a highly optimized version of logistic regression implementation, which supports multiclass classification tasks, \cite{16}.

\subsection{Random Forest Classifier}
Random forest classifier, \cite{16}, is an ensemble method that trains multiple decision trees in parallel with bootstrapping which is followed by aggregation. The bootstrapping indicates that the different individual decision trees are trained concurrently on multiple subsets of the training dataset using different subsets of the available features. The bootstrapping will ensure that every individual decision tree is unique, and that reduces the overall variance of the random forest classifier. For the final decision step, the random forest classifier aggregates the decisions of all the individual trees; then, the classifier will exhibit good generalization. Random forest classifier tends to outperform most other classification methods in accuracy without having issues of overfitting. The random forest classifier doesn't require the feature scaling process. Even though a random forest classifier is harder to interpret, it's easier to tune the hyperparameter when we compare it to a decision tree classifier. The general figure of random forest is represented in Figure \ref{fig:rf}.

\subsection{Multi layer perceptron (MLP)}
\begin{figure}[ht]
\centering
\captionsetup{justification=centering}
  \includegraphics[scale=0.35]{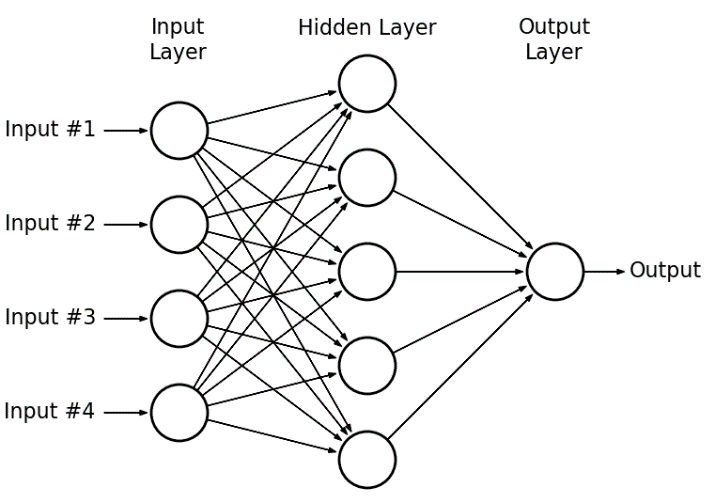}
  \caption{Multi layer perceptron}
  \label{fig:mlp}
\end{figure}

MLP can be viewed as a supplement of feed-forward neural network, \cite{17}, which consists of three different types of layers which are; the input, output, and hidden layers. The input layer received the input signal for processing. The prediction and classification are performed by the output layer. A number of hidden layers that are placed between the input and the output layer are the core engine of the MLP. Like a feed-forward network, the data flows in a forward direction from the input to the output layer. The neurons in the MLP are trained with the backpropagation learning algorithm. MLPs can solve problems that are not linearly separable. The major use cases of Multi-layer perceptron are pattern recognition, classification, approximation, and prediction. The computations in MPL take place at every neuron in the output and hidden layer \cite{18}. Figure \ref{fig:mlp} shows the input, hidden and output lyers of a MLP.

\subsection{Gradient Boosting Classifier}
Gradient Boosting is a machine learning algorithm, that's used for both classification and regression problems. This classifier works on the idea that multiple weak learners can collaborate together and make a more accurate predictor. Gradient boosting classifier works by building simpler and weak prediction models sequentially where each model will try to predict the error left from the previous model. That's why this algorithm tends to over-fit quickly \cite{25}.

\subsection{Decision Tree Classifier}
A decision tree is a versatile machine learning algorithm used for classification and regression tasks. It operates by splitting a dataset into smaller, manageable subsets while simultaneously developing a tree-like model of decisions. Each node in the tree represents a decision point, leading to branches and ultimately to leaf nodes that signify outcomes. The simplicity and visual interpretability of decision trees make them easily comprehensible, ideal for practical decision-making scenarios. However, they can be prone to overfitting, especially with complex datasets. To mitigate this, techniques like pruning are employed. Decision trees also lay the groundwork for advanced ensemble methods such as Random Forests and Gradient Boosted Trees, enhancing predictive performance and robustness \cite{24}\cite{26}.

\subsection{Other methods}
In addition to the method described previously, we used plenty of other methods to build and train models using the input dataset. These methods are the K-Neighbors classifier \cite{27},  ADA boosting classifier \cite{28}, Bagging classifier \cite{29}, Stacking classifier \cite{30} and Naive Bayes classifiers (Bernoulli and Multinomial) \cite{31}.
All the nine mentioned methods went through the same process from model evaluation and finally selecting the best model.

\section{Results and discussion}
Our research utilized a dataset of approximately 148.3K URLs, with one-third categorized as spam and the remaining as non-spam. For model evaluation and training, the data was partitioned with \SI{20}{\percent} allocated for testing and \SI{80}{\percent} for training.
\begin{figure*}[!htb]
    \centering
    \begin{subfigure}[b]{0.32\linewidth}
        \centering
        \includegraphics[width=\textwidth]{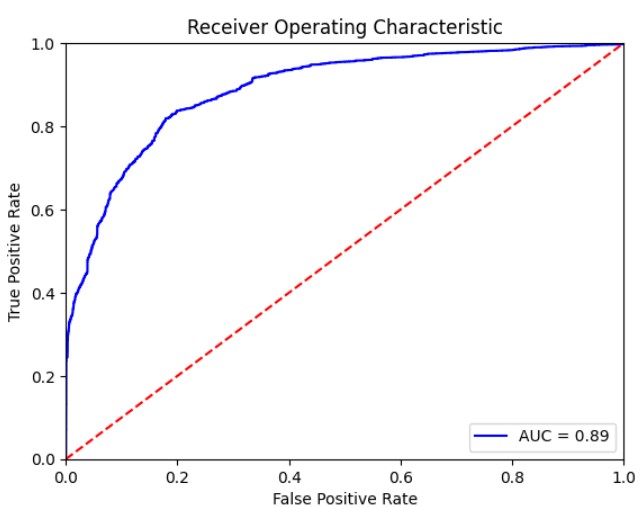}
        \caption*{Logistic Regression }
    \end{subfigure}
    \hfill
    \begin{subfigure}[b]{0.32\linewidth}
        \centering
        \includegraphics[width=\textwidth]{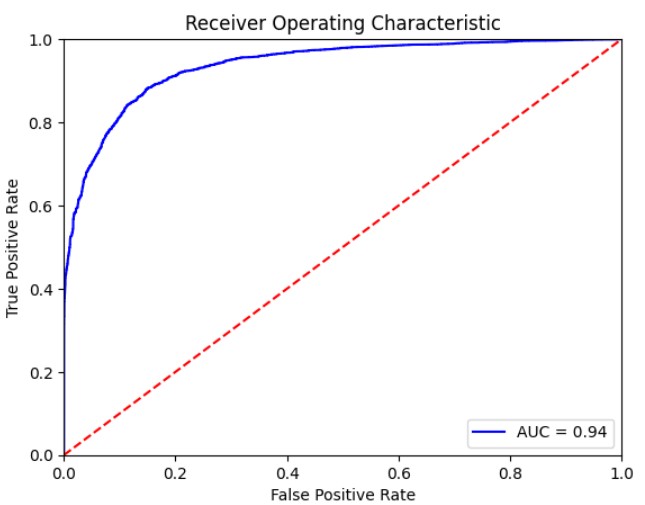}
        \caption*{Multi-layer Perceptron }
    \end{subfigure}
    \hfill
    \begin{subfigure}[b]{0.32\linewidth}
        \centering
        \includegraphics[width=\textwidth]{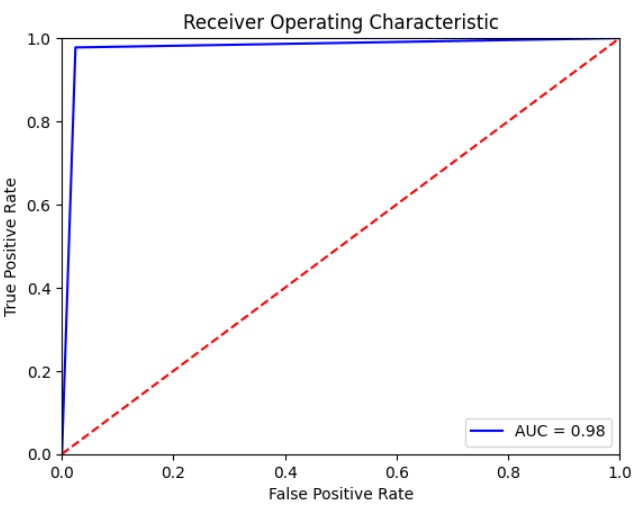}
        \caption*{KNeighbours }
    \end{subfigure}
    
    \vspace{10pt}
    
    \begin{subfigure}[b]{0.32\linewidth}
        \centering
        \includegraphics[width=\textwidth]{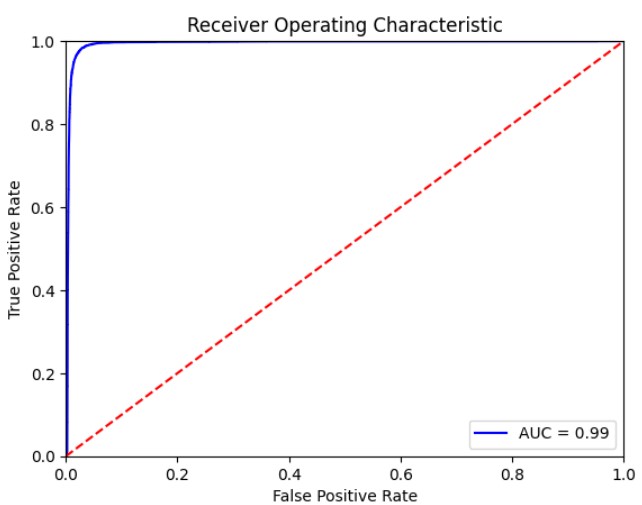}
        \caption*{Gradient Boosting }
    \end{subfigure}
    \hfill
    \begin{subfigure}[b]{0.32\linewidth}
        \centering
        \includegraphics[width=\textwidth]{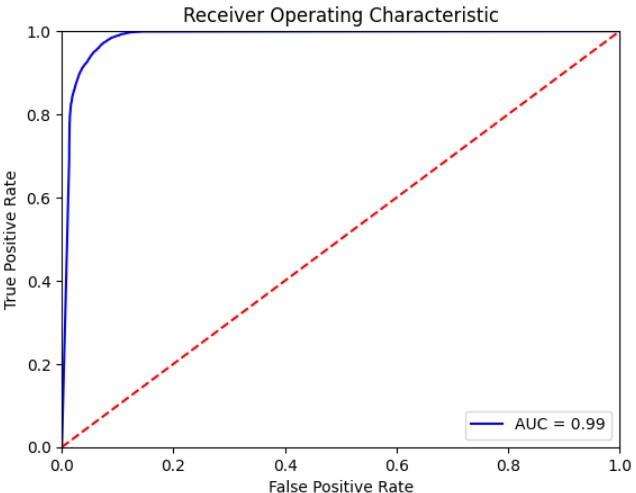}
        \caption*{Decision Tree }
    \end{subfigure}
    \hfill
    \begin{subfigure}[b]{0.32\linewidth}
        \centering
        \includegraphics[width=\textwidth]{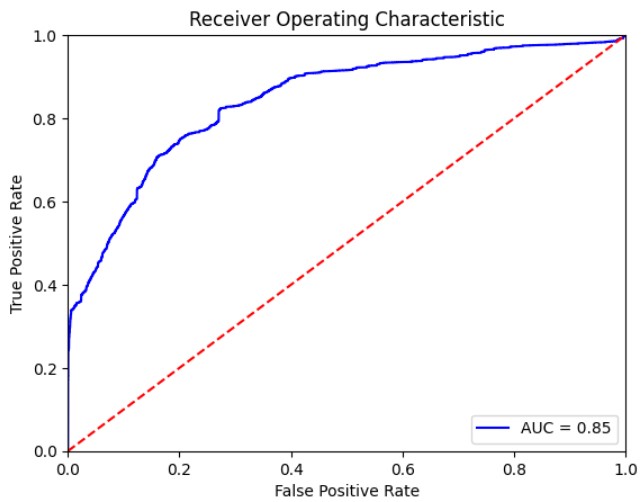}
        \caption*{Multinomial }
    \end{subfigure}
    
    \vspace{10pt}
    
    \begin{subfigure}[b]{0.32\linewidth}
        \centering
        \includegraphics[width=\textwidth]{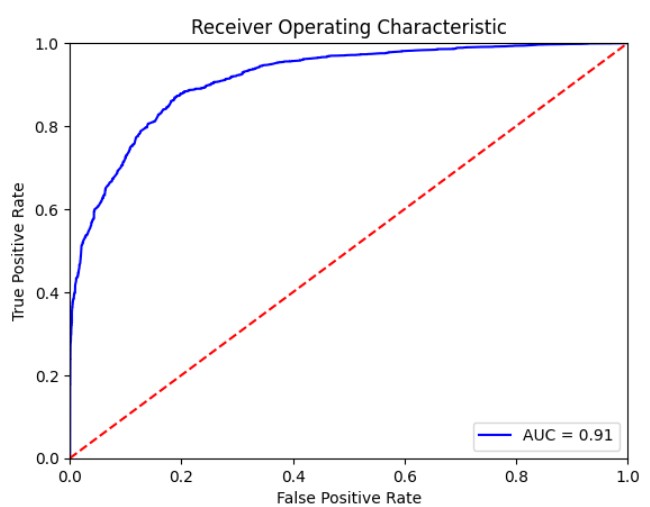}
        \caption*{Ada Boost }
    \end{subfigure}
    \hfill
    \begin{subfigure}[b]{0.32\linewidth}
        \centering
        \includegraphics[width=\textwidth]{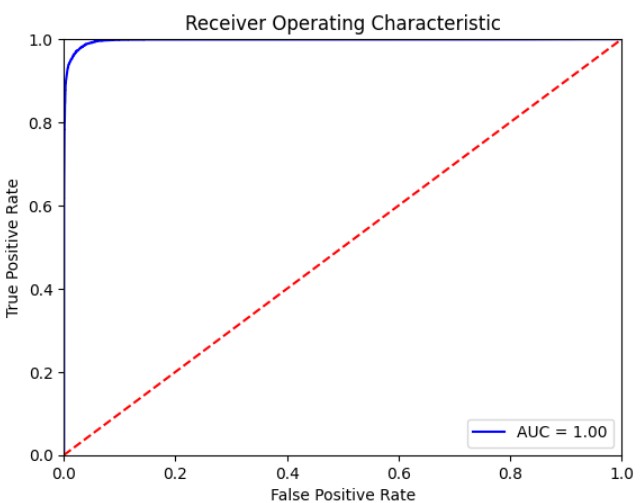}
        \caption*{Random Forest }
    \end{subfigure}
    \hfill
    \begin{subfigure}[b]{0.32\linewidth}
        \centering
        \includegraphics[width=\textwidth]{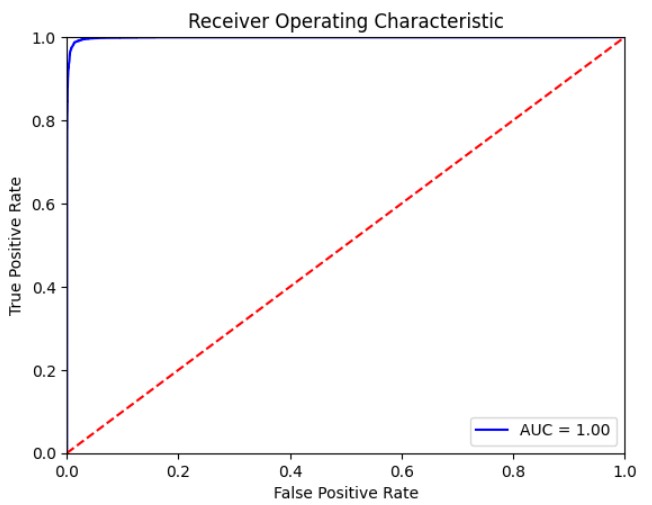}
        \caption*{Bagging }
    \end{subfigure}

       \vspace{10pt}
    
    \begin{subfigure}[b]{0.32\linewidth}
        \centering
        \includegraphics[width=\textwidth]{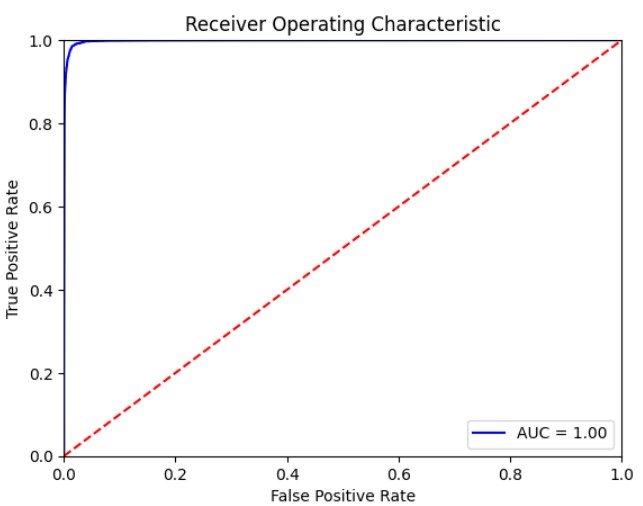}
        \caption*{Stacking }
    \end{subfigure}
    \hfill
    \begin{subfigure}[b]{0.32\linewidth}
        \centering
        \includegraphics[width=\textwidth]{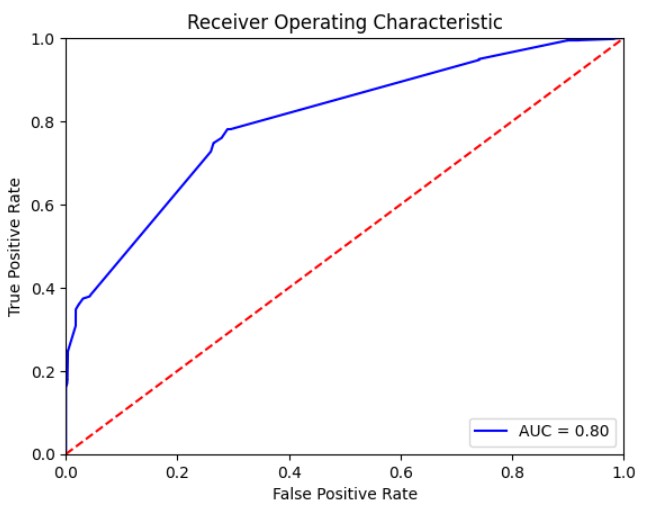}
        \caption*{Bernoulli }
    \end{subfigure}
    \hfill
    \begin{subfigure}[b]{0.32\linewidth}
        \centering
        \includegraphics[width=\textwidth]{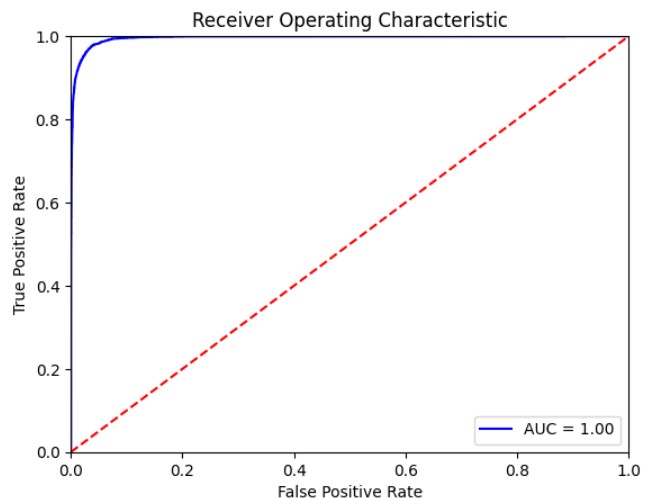}
        \caption*{LightGBM }
    \end{subfigure}
    \caption{ROC and AUC curves for various classifiers.}
\end{figure*}

In optimizing the classifiers, hyperparameters were fine-tuned through a random search methodology, balancing computational efficiency with the likelihood of securing optimal hyperparameter combinations.

\subsection{Comparison between all the proposed ML models}
\begin{figure*}[htp]
    \centering
    \begin{subfigure}[b]{0.3\linewidth}
        \centering
        \includegraphics[width=\textwidth]{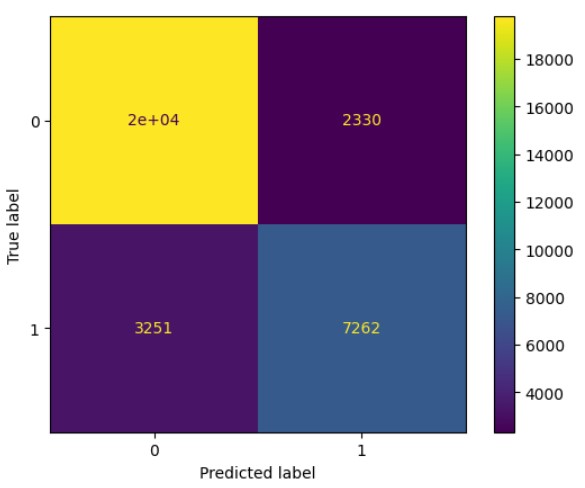}
        \caption*{Logistic Regression}
    \end{subfigure}
    \hfill
    \begin{subfigure}[b]{0.3\linewidth}
        \centering
        \includegraphics[width=\textwidth]{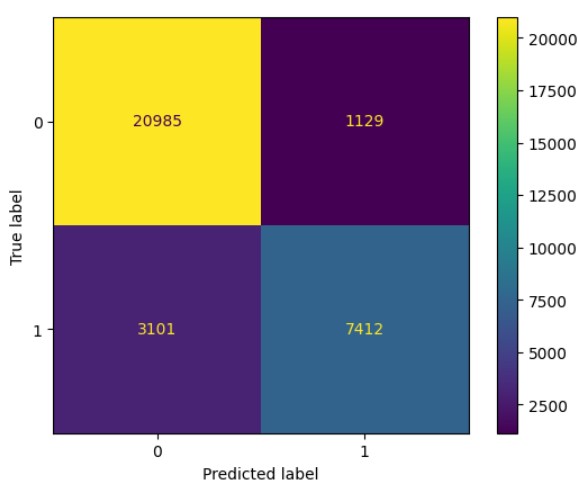}
        \caption*{MLP Classifier(NueralNetwork)}
    \end{subfigure}
    \hfill
    \begin{subfigure}[b]{0.3\linewidth}
        \centering
        \includegraphics[width=\textwidth]{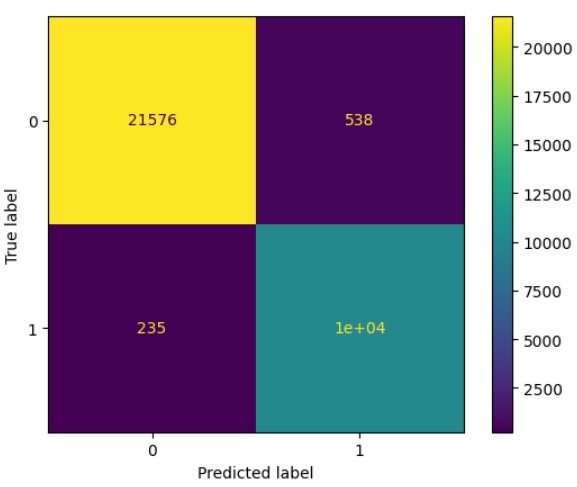}
        \caption*{KNN}
    \end{subfigure}
    \\ % line break
    \begin{subfigure}[b]{0.3\linewidth}
        \centering
        \includegraphics[width=\textwidth]{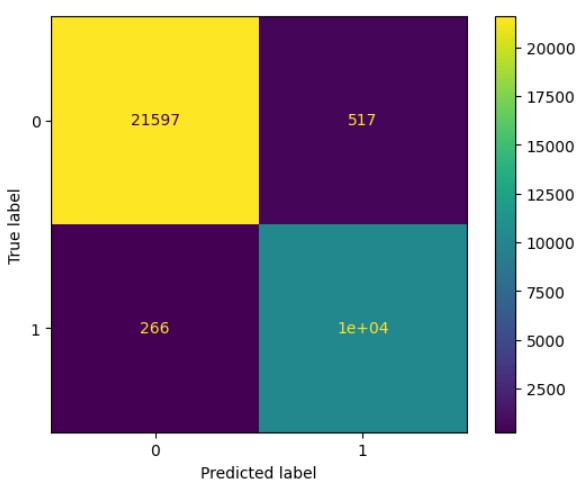}
        \caption*{Gradient Boosting}
    \end{subfigure}
    \hfill
    \begin{subfigure}[b]{0.3\linewidth}
        \centering
        \includegraphics[width=\textwidth]{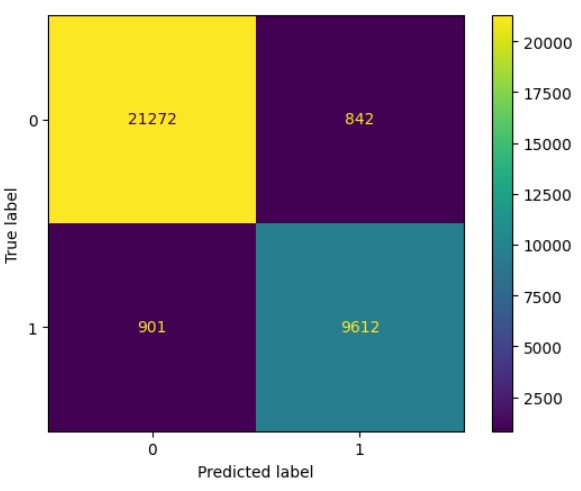}
        \caption*{Decision Tree}
    \end{subfigure}
    \hfill
    \begin{subfigure}[b]{0.3\linewidth}
        \centering
        \includegraphics[width=\textwidth]{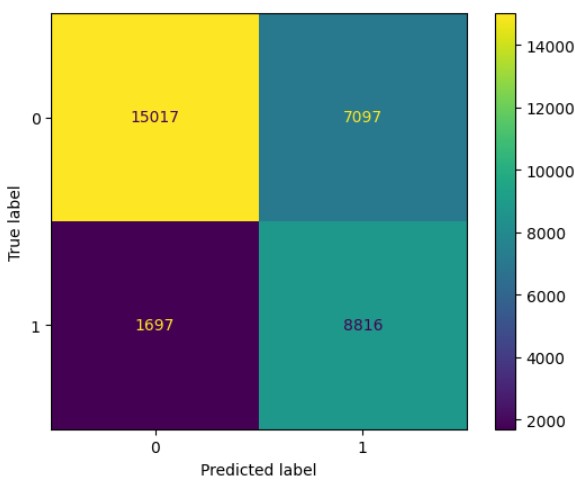}
        \caption*{MultinomialNB}
    \end{subfigure}
    \\ % line break
    \begin{subfigure}[b]{0.3\linewidth}
        \centering
        \includegraphics[width=\textwidth]{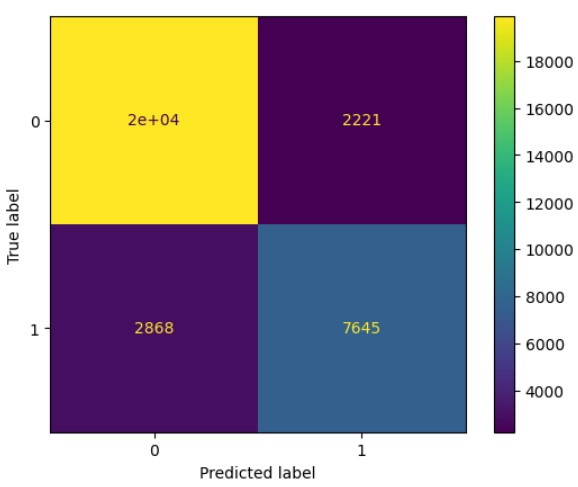}
        \caption*{Ada Boost}
    \end{subfigure}
    \hfill
    \begin{subfigure}[b]{0.3\linewidth}
        \centering
        \includegraphics[width=\textwidth]{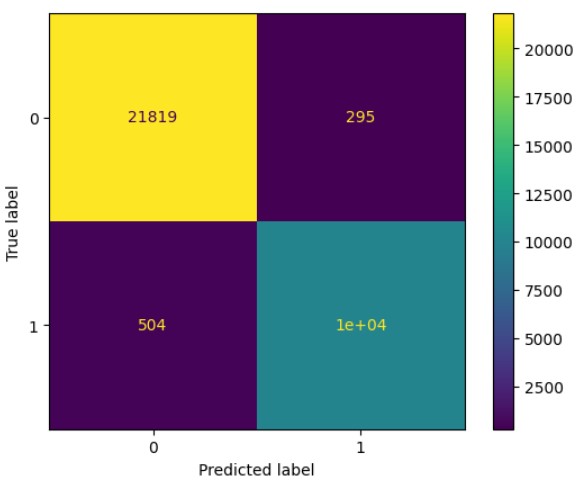}
        \caption*{Random Forest}
    \end{subfigure}
    \hfill
    \begin{subfigure}[b]{0.3\linewidth}
        \centering
        \includegraphics[width=\textwidth]{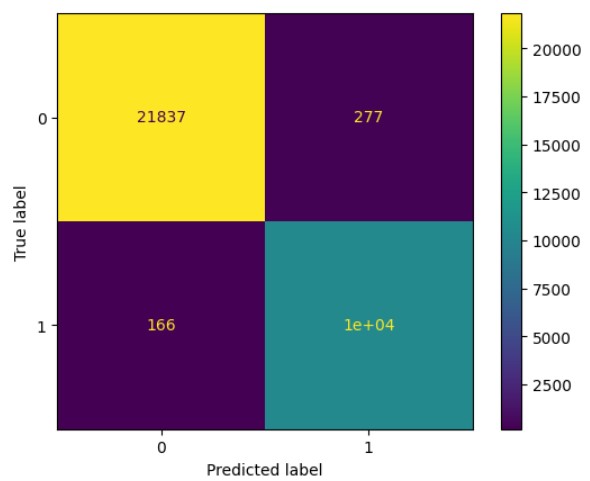}
        \caption*{Bagging}
    \end{subfigure}
     \\ % line break
    \begin{subfigure}[b]{0.3\linewidth}
        \centering;l
        \includegraphics[width=\textwidth]{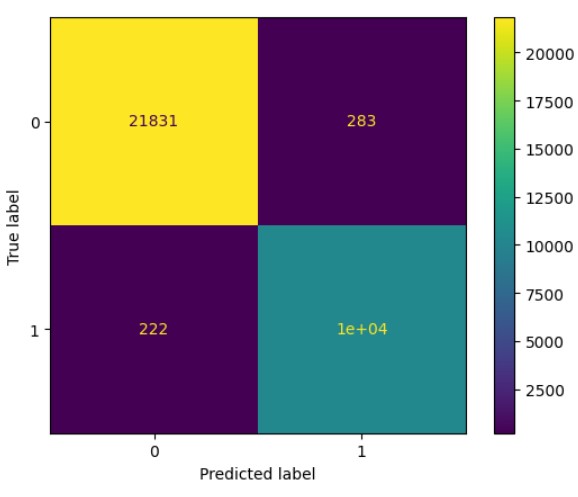}
        \caption*{Stacking}
    \end{subfigure}
    \hfill
    \begin{subfigure}[b]{0.3\linewidth}
        \centering
        \includegraphics[width=\textwidth]{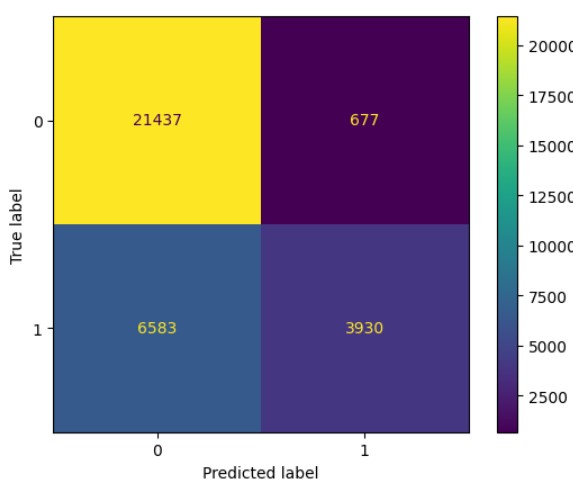}
        \caption*{BernoulliNB}
    \end{subfigure}
    \hfill
    \begin{subfigure}[b]{0.3\linewidth}
        \centering
        \includegraphics[width=\textwidth]{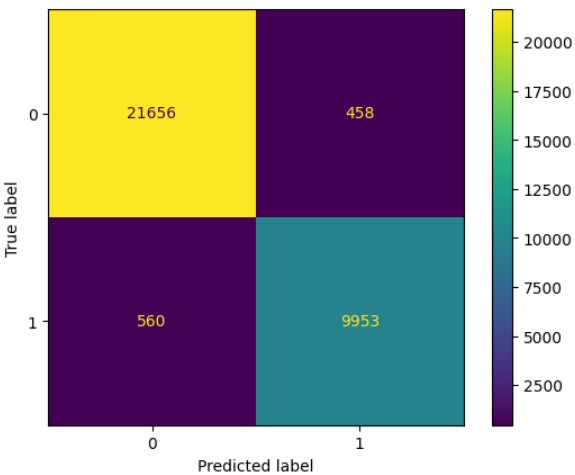}
        \caption*{LightGBM}
    \end{subfigure}
    \caption{Confusion matrices of various classifiers.}
\label{figure:confusion-matricies}
\end{figure*}

For assessing classifier performance, the AUC and ROC curves were employed, As Shown in Figure \ref{figure:confusion-matricies}, The Random Forest, Bagging, and Stacking classifiers each achieved a perfect  AUC score of 1. Conversely, the Multinomial classifier and the Bernoulli classifier registered AUC scores of 0.85 and 0.8, respectively. Further insights were derived from the confusion matrices, As Shown in Figure 8, which underscored Bagging 

\noindent as the predominant classifier in terms of precision and recall, closely followed by Stacking.

Figure 8 serves as a pivotal component of our analysis, providing a comprehensive view of twelve confusion matrices generated by various classification methods. These matrices allow for a thorough evaluation of classifier performance based on their false positive and false negative rates. Among these classifiers, the Bagging Classifier emerges as the top performer, displaying a commendable equilibrium between false positives (166) and false negatives (277). This balance underscores the classifier's effectiveness in correctly classifying instances while minimizing the occurrence of misclassifications.

In contrast, the Bernoulli Naive Bayes (BernoulliNB) Classifier demonstrates the least favorable performance, with an alarmingly high
\begin{table*}[!htbp]
\caption{Comparison results between ML models for spam detection}
\centering
\captionsetup{justification=centering}
\small
\begin{tabularx}{\textwidth}{lXXXXXXX}
\toprule
Classifier & Acc. & 10 K-fold & Prec. & Recall & F1 & R2 \\
\midrule
LogisticRegression                & 82.89\% & 82.61\% & 75.71\% & 69.08\% & 72.24\% & 0.22 \\
KNeighborsClassifier              & 97.63\% & 96.63\% & 95.03\% & 97.76\% & 96.38\% & 0.89 \\
AdaBoostClassifier                & 84.40\% & 84.33\% & 77.49\% & 72.72\% & 75.03\% & 0.29 \\
MultinominalNB                    & 73.05\% & 72.90\% & 55.40\% & 83.86\% & 66.72\% & -0.23 \\
BernoulliNB                       & 77.75\% & 78.19\% & 85.30\% & 37.38\% & 51.98\% & -0.02 \\
RandomForestClassifier            & 97.55\% & 96.87\% & 97.14\% & 95.21\% & 96.16\% & 0.89 \\
GradientBoostingClassifier        & 97.60\% & 96.64\% & 95.20\% & 97.47\% & 96.32\% & 0.89 \\
MLPClassifier(NueralNetwork)      & 87.04\% & 87.15\% & 86.78\% & 70.50\% & 77.80\% & 0.41 \\
BaggingClassifier                 & 98.64\% & 97.93\% & 97.39\% & 98.42\% & 97.90\% & 0.94 \\
StackingClassifier                & 98.45\% & 97.82\% & 97.32\% & 97.89\% & 97.61\% & 0.93 \\
DecisionTreeClassifier            & 94.66\% & 94.05\% & 91.95\% & 91.43\% & 91.69\% & 0.76 \\
lightgbmClassifier                & 96.88\% & 96.09\% & 95.60\% & 94.67\% & 95.13\% & 0.86 \\
\bottomrule
\label{table:ML cmoparison}
\end{tabularx}
\end{table*}

\noindent false positive count of 6583 and an elevated false negative count of 677. This classifier faces notable challenges in achieving accurate classifications, resulting in a substantial number of both false positives and false negatives. Similarly, the Logistic

\noindent Regression Classifier exhibits suboptimal performance, with a relatively high false positive count of 3251 and a substantial false negative count of 2330. These findings emphasize the crucial role of careful classifier selection and potential fine-tuning in achieving robust classification results.

The Bagging Classifier, AS Shown in \ref{table:ML cmoparison} was particularly noteworthy with an accuracy of \SI{98.64}{\percent} and a 10 K-fold validation score of \SI{97.93}{\percent}. Its performance metrics, including precision, recall, and F1 score, further accentuated its dominance. The Stacking Classifier, though impressive, was slightly behind with an accuracy of \SI{98.45}{\percent} and a 10 K-fold validation score of \SI{97.82}{\percent}. Despite the perfect AUC score for the Random Forest, its accuracy was \SI{97.55}{\percent} with a 10 K-fold score of \SI{96.87}{\percent} On the flip side, the MultinomialNB and BernoulliNB classifiers lagged, with respective accuracies of \SI{73.05}{\percent} and \SI{77.75}{\percent}.

\subsection{Comparison with the state-of-the-art methods}
In this section, we compare the performance of the best-tested ML model against the most related models in the literature. As shown in Table \ref{table:spam-detection}, many studies on spam URL detection using machine learning have highlighted significant advancements. The results presented in \cite{32}\cite{33}\cite{34} demonstrated high effectiveness with models like DistilBERT and Random Forest, achieving accuracies ranging from 93.77\% to 97.39\%. In our study, we outperformed these benchmarks, achieving an accuracy of 98.64\% using a Bagging Classifier. These findings underscore the evolving efficacy of diverse machine learning techniques in addressing spam URLs, emphasizing the imperative for ongoing innovation in cybersecurity.

In summary, while many classifiers demonstrated commendable efficacy in distinguishing between spam and non-spam URLs, the Bagging classifier, as indicated by the presented metrics, emerge as top contenders for practical applications.

\begin{table*}[!ht]
\centering
\caption{Comparison between our model and the state-of-the-art models for spam detection}
\captionsetup{justification=centering}
\small
\begin{tabularx}{\textwidth}{Xccc}
\toprule
Article reference & Year & Classifier Used & Accuracy (10 k fold) \\
\midrule
\cite{32} & 2022 & DistilBERT & 95.02\% \\ 

\cite{33} & 2022 & Random Forest & 97.39\% \\ 

\cite{34} & 2022 & Random Forest & 93.77\% \\ 

(Our study) & 2023 & Bagging Classifier & 98.64\% \\ 
\bottomrule
\label{table:spam-detection}
\end{tabularx}
\end{table*}

\section{Conclusion}

To sum up the work done and discussed in this paper, we utilized a dataset that classifies URLs as spam or not spam, analyzed the data, and extracted multiple features. Then we trained various machine learning models using this dataset. For each model, we tuned the hyperparameters and cross-validated the results. The outcomes showed several models with accuracy higher than 90%. After evaluating all the models, the Bagging classifier emerged as the top performer, boasting the highest accuracy of 98.64%.
\newline
For a future work, we would like to train the data sets on more models like deep learning models, and we also would like to extract more features from the website of the URL itself such as the body size of the web page and features related to the script used as well.

\bibliographystyle{IEEEtran} % We choose the "plain" reference style
\bibliography{refs} % Entries are in the refs.bib file

\begin{thebibliography}{10}
\providecommand{\url}[1]{#1}
\csname url@samestyle\endcsname
\providecommand{\newblock}{\relax}
\providecommand{\bibinfo}[2]{#2}
\providecommand{\BIBentrySTDinterwordspacing}{\spaceskip=0pt\relax}
\providecommand{\BIBentryALTinterwordstretchfactor}{4}
\providecommand{\BIBentryALTinterwordspacing}{\spaceskip=\fontdimen2\font plus
\BIBentryALTinterwordstretchfactor\fontdimen3\font minus
  \fontdimen4\font\relax}
\providecommand{\BIBforeignlanguage}[2]{{%
\expandafter\ifx\csname l@#1\endcsname\relax
\typeout{** WARNING: IEEEtran.bst: No hyphenation pattern has been}%
\typeout{** loaded for the language `#1'. Using the pattern for}%
\typeout{** the default language instead.}%
\else
\language=\csname l@#1\endcsname
\fi
#2}}
\providecommand{\BIBdecl}{\relax}
\BIBdecl

\bibitem{1}
W.~Zhongtao, P.~Xin, W.~Yuling, L.~Yaohua, H.~Li, and C.~Biao, ``Analysis on
  the characteristics of url spam,'' vol.~1, 03 2012.

\bibitem{2}
A.~Arram, H.~Mousa, and A.~Zainal, ``Spam detection using hybrid artificial
  neural network and genetic algorithm,'' in \emph{2013 13th International
  Conference on Intellient Systems Design and Applications}, 2013, pp.
  336--340.

\bibitem{3}
F.~Vanhoenshoven, G.~N{\'a}poles, R.~Falcon, K.~Vanhoof, and M.~K{\"o}ppen,
  ``Detecting malicious urls using machine learning techniques,'' in \emph{2016
  IEEE Symposium Series on Computational Intelligence (SSCI)}.\hskip 1em plus
  0.5em minus 0.4em\relax IEEE, 2016, pp. 1--8.

\bibitem{4}
M.~I. Jordan and T.~M. Mitchell, ``Machine learning: Trends, perspectives, and
  prospects,'' \emph{Science}, vol. 349, no. 6245, pp. 255--260, 2015.

\bibitem{22}
A.~Arram, M.~Ayob, and A.~Sulaiman, ``Hybrid bird mating optimizer with
  single-based algorithms for combinatorial optimization problems,'' \emph{IEEE
  Access}, vol.~9, pp. 115\,972--115\,989, 2021.

\bibitem{5}
M.~Anjaneyulu, B.~Madhuravani, and P.~Devika, ``Detection of malicious websites
  using machine learning approach and web vulnerability scanner,'' in \emph{AIP
  Conference Proceedings}, vol. 2492, no.~1.\hskip 1em plus 0.5em minus
  0.4em\relax AIP Publishing, 2023.

\bibitem{6}
M.~A.~A. Albadr, M.~Ayob, S.~Tiun, F.~T. AL-Dhief, A.~Arram, and S.~Khalaf,
  ``Breast cancer diagnosis using the fast learning network algorithm,''
  \emph{Frontiers in Oncology}, vol.~13, p. 1150840, 2023.

\bibitem{7}
A.~Sulaiman, K.~Omar, M.~F. Nasrudin, and A.~Arram, ``Length independent writer
  identification based on the fusion of deep and hand-crafted descriptors,''
  \emph{IEEE Access}, vol.~7, pp. 91\,772--91\,784, 2019.

\bibitem{8}
Z.~Gyongyi and H.~Garcia-Molina, ``Web spam taxonomy,'' in \emph{First
  international workshop on adversarial information retrieval on the web
  (AIRWeb 2005)}, 2005.

\bibitem{21}
A.~Arram and M.~Ayob, ``A novel multi-parent order crossover in genetic
  algorithm for combinatorial optimization problems,'' \emph{Computers \&
  Industrial Engineering}, vol. 133, pp. 267--274, 2019.

\bibitem{18}
S.~Abirami and P.~Chitra, ``Energy-efficient edge based real-time healthcare
  support system,'' in \emph{Advances in Computers}.\hskip 1em plus 0.5em minus
  0.4em\relax Elsevier, 2020, vol. 117, no.~1, pp. 339--368.

\bibitem{20}
K.~Murat, A.~{\.I}sa, and M.~A.~S. AL-HAYAN{\.I}, ``Classification of malicious
  urls using naive bayes and genetic algorithm,'' \emph{Sakarya University
  Journal of Computer and Information Sciences}, vol.~6, no.~2, pp. 80--90,
  2023.

\bibitem{9}
A.~Ntoulas, M.~Najork, M.~Manasse, and D.~Fetterly, ``Detecting spam web pages
  through content analysis,'' in \emph{Proceedings of the 15th international
  conference on World Wide Web}, 2006, pp. 83--92.

\bibitem{10}
M.~Egele, C.~Kolbitsch, and C.~Platzer, ``Removing web spam links from search
  engine results,'' \emph{Journal in Computer Virology}, vol.~7, no.~1, pp.
  51--62, 2011.

\bibitem{11}
B.~E. Boser, I.~M. Guyon, and V.~N. Vapnik, ``A training algorithm for optimal
  margin classifiers,'' in \emph{Proceedings of the fifth annual workshop on
  Computational learning theory}, 1992, pp. 144--152.

\bibitem{12}
T.~Joachims, ``Text categorization with support vector machines: Learning with
  many relevant features,'' in \emph{European conference on machine
  learning}.\hskip 1em plus 0.5em minus 0.4em\relax Springer, 1998, pp.
  137--142.

\bibitem{23}
D.~Albashish, H.~M. Mustafa, R.~A. Khurma, B.~Hasan, S.~Bani-Ahmad,
  A.~Abdullah, and A.~Arram, ``Enhanced meta-heuristic methods for industrial
  winding process modelling,'' \emph{Expert Systems}, p. e13438.

\bibitem{13}
W.~Tsai, C.~Chang, M.~Lin, S.~Chien, H.~Sun, and M.~Hsieh, ``Adsorption of acid
  dye onto activated carbons prepared from agricultural waste bagasse by zncl2
  activation,'' \emph{Chemosphere}, vol.~45, no.~1, pp. 51--58, 2001.

\bibitem{14}
\BIBentryALTinterwordspacing
S.~Bansal, ``Spam urls classification dataset,'' 2021. [Online]. Available:
  \url{https://www.kaggle.com/shivamb/spam-url-prediction}
\BIBentrySTDinterwordspacing

\bibitem{15}
\BIBentryALTinterwordspacing
T.~pudding. (2021) The pudding. [Online]. Available:
  \url{https://pudding.cool/}
\BIBentrySTDinterwordspacing

\bibitem{16}
A.~Subasi, \emph{Practical Machine Learning for Data Analysis Using Python}, 06
  2020.

\bibitem{17}
S.~Misra, H.~Li, and J.~He, \emph{Machine learning for subsurface
  characterization}.\hskip 1em plus 0.5em minus 0.4em\relax Gulf Professional
  Publishing, 2019.

\bibitem{25}
L.~Lusa \emph{et~al.}, ``Gradient boosting for high-dimensional prediction of
  rare events,'' \emph{Computational Statistics \& Data Analysis}, vol. 113,
  pp. 19--37, 2017.

\bibitem{24}
A.~Arram, M.~Ayob, M.~A.~A. Albadr, A.~Sulaiman, and D.~Albashish, ``Credit
  card score prediction using machine learning models: A new dataset,''
  \emph{arXiv preprint arXiv:2310.02956}, 2023.

\bibitem{26}
A.~J. Myles, R.~N. Feudale, Y.~Liu, N.~A. Woody, and S.~D. Brown, ``An
  introduction to decision tree modeling,'' \emph{Journal of Chemometrics: A
  Journal of the Chemometrics Society}, vol.~18, no.~6, pp. 275--285, 2004.

\bibitem{27}
P.~Viswanath and T.~H. Sarma, ``An improvement to k-nearest neighbor
  classifier,'' in \emph{2011 IEEE Recent Advances in Intelligent Computational
  Systems}.\hskip 1em plus 0.5em minus 0.4em\relax IEEE, 2011, pp. 227--231.

\bibitem{28}
H.~Schwenk and Y.~Bengio, ``Boosting neural networks,'' \emph{Neural
  computation}, vol.~12, no.~8, pp. 1869--1887, 2000.

\bibitem{29}
M.~Skurichina and R.~P. Duin, ``Bagging for linear classifiers,'' \emph{Pattern
  Recognition}, vol.~31, no.~7, pp. 909--930, 1998.

\bibitem{30}
G.~Sakkis, I.~Androutsopoulos, G.~Paliouras, V.~Karkaletsis, C.~D. Spyropoulos,
  and P.~Stamatopoulos, ``Stacking classifiers for anti-spam filtering of
  e-mail,'' \emph{arXiv preprint cs/0106040}, 2001.

\bibitem{31}
T.~Bayes, ``Naive bayes classifier,'' \emph{Article Sources and Contributors},
  pp. 1--9, 1968.

\bibitem{32}
S.~Kotni and D.~L.~S. Chandrasekhar~Potala, ``Spam detection using deep
  learning models.''

\bibitem{33}
A.~K. Jilani and J.~Sultana, ``A random forest based approach to classify spam
  urls data,'' in \emph{2022 ASU International Conference in Emerging
  Technologies for Sustainability and Intelligent Systems (ICETSIS)}.\hskip 1em
  plus 0.5em minus 0.4em\relax IEEE, 2022, pp. 268--272.

\bibitem{34}
M.~YILDIRIM, ``Using and comparing machine learning techniques for automatic
  detection of spam website urls,'' \emph{NATURENGS}, vol.~3, no.~1, pp.
  33--41, 2022.

\end{thebibliography}

\end{document}